
\newbox\leftpage \newdimen\fullhsize \newdimen\hstitle \newdimen\hsbody
\tolerance=1000\hfuzz=2pt
\def\bigans{b }
\def\answ{b }
%
\ifx\answ\bigans\message{(This will come out unreduced.}
\hsbody=\hsize \hstitle=\hsize 
\else\def\apans{l }\message{ lyman or hepl (l/h) (lowercase]) ? }
\read-1 to \apansw\message{(This will be reduced.}
\let\lr=L
\voffset=-.31truein\vsize=7truein
\hstitle=8truein\hsbody=4.75truein\fullhsize=10truein\hsize=\hsbody
\ifx\apansw\apans\special{ps: landscape}\hoffset=-.59truein
  \else\hoffset=.05truein\fi
\output={\ifnum\pageno=0 
  \shipout\vbox{\hbox to \fullhsize{\hfill\pagebody\hfill}}\advancepageno
  \else
  \almostshipout{\leftline{\vbox{\pagebody\makefootline}}}\advancepageno
  \fi}
\def\almostshipout#1{\if L\lr \count1=1
      \global\setbox\leftpage=#1 \global\let\lr=R
  \else \count1=2
    \shipout\vbox{\ifx\apansw\apans\special{ps: landscape}\fi 
      \hbox to\fullhsize{\box\leftpage\hfil#1}}  \global\let\lr=L\fi}
\fi
%
\catcode`\@=11 
\newcount\yearltd\yearltd=\year\advance\yearltd by -1900

%
%
\def\draftmode{\def\draftdate{{\rm preliminary draft:
\number\month/\number\day/\number\yearltd\ \ \hourmin}}%
\headline={\hfil\draftdate}\writelabels\baselineskip=20pt plus 2pt minus 2pt
{\count255=\time\divide\count255 by 60 \xdef\hourmin{\number\count255}
        \multiply\count255 by-60\advance\count255 by\time
   \xdef\hourmin{\hourmin:\ifnum\count255<10 0\fi\the\count255}}}

\def\nolabels{\def\eqnlabel##1{}\def\eqlabel##1{}\def\reflabel##1{}}
\def\writelabels{\def\eqnlabel##1{%
{\escapechar=` \hfill\rlap{\hskip.09in\string##1}}}%
\def\eqlabel##1{{\escapechar=` \rlap{\hskip.09in\string##1}}}%
\def\reflabel##1{\noexpand\llap{\string\string\string##1\hskip.31in}}}
\nolabels
%
\global\newcount\secno \global\secno=-1
\global\newcount\meqno \global\meqno=1
\def\newsec#1{\global\advance\secno by1\message{(\the\secno. #1)}
\xdef\secsym{\the\secno.}\global\meqno=1
\bigbreak\bigskip
\noindent{\bf\the\secno. #1}\par\nobreak\medskip\nobreak}
\xdef\secsym{}
\def\appendix#1#2{\global\meqno=1\xdef\secsym{\hbox{#1.}}\bigbreak\bigskip
\noindent{\bf Appendix #1. #2}\par\nobreak\medskip\nobreak}
%
%
\def\eqnn#1{\xdef #1{(\secsym\the\meqno)}%
\global\advance\meqno by1\eqnlabel#1}
\def\eqna#1{\xdef #1##1{\hbox{$(\secsym\the\meqno##1)$}}%
\global\advance\meqno by1\eqnlabel{#1$\{\}$}}
\def\eqn#1#2{\xdef #1{(\secsym\the\meqno)}\global\advance\meqno by1%
$$#2\eqno#1\eqlabel#1$$}
%
\newskip\footskip\footskip14pt plus 1pt minus 1pt 
\def\f@@t{\baselineskip\footskip\bgroup\aftergroup\@foot\let\next}
\setbox\strutbox=\hbox{\vrule height9.5pt depth4.5pt width0pt}
\global\newcount\ftno \global\ftno=0
\def\foot{\global\advance\ftno by1\footnote{$^{\the\ftno}$}}
%
%
\global\newcount\refno \global\refno=1
\newwrite\rfile
\def\ref{\nref}
\def\nref#1{\xdef#1{[\the\refno]}\ifnum\refno=1\immediate
\openout\rfile=refs.tmp\fi\global\advance\refno by1\chardef\wfile=\rfile
\immediate\write\rfile{\noexpand\item{#1\ }\reflabel{#1}\pctsign}\findarg}
\def\findarg#1#{\begingroup\obeylines\newlinechar=`\^^M\pass@rg}
{\obeylines\gdef\pass@rg#1{\writ@line\relax #1^^M\hbox{}^^M}%
\gdef\writ@line#1^^M{\expandafter\toks0\expandafter{\striprel@x #1}%
\edef\next{\the\toks0}\ifx\next\em@rk\let\next=\endgroup\else\ifx\next\empty%
\else\immediate\write\wfile{\the\toks0}\fi\let\next=\writ@line\fi\next\relax}}
\def\striprel@x#1{} \def\em@rk{\hbox{}} {\catcode`\%=12\xdef\pctsign{
\def\semi{;\hfil\break}
\def\addref#1{\immediate\write\rfile{\noexpand\item{}#1}} 
\def\listrefs{\immediate\closeout\rfile
\baselineskip=14pt\centerline{{\bf References}}\bigskip{\frenchspacing%
\escapechar=` \input refs.tmp\vfill\eject}\nonfrenchspacing}
\def\startrefs#1{\immediate\openout\rfile=refs.tmp\refno=#1}
\def\figures{\centerline{{\bf Figure Captions}}\medskip\parindent=40pt}
\def\fig#1#2{\medskip\item{Fig.~#1:  }#2}
\catcode`\@=12 
%
%
\def\noblackbox{\overfullrule=0pt}
\hyphenation{anom-aly anom-alies coun-ter-term coun-ter-terms}
\def\inv{^{\raise.15ex\hbox{${\scriptscriptstyle -}$}\kern-.05em 1}}
\def\dup{^{\vphantom{1}}}
\def\Dsl{\,\raise.15ex\hbox{/}\mkern-13.5mu D} 
\def\dsl{\raise.15ex\hbox{/}\kern-.57em\partial}
\def\del{\partial}
\def\Psl{\dsl}
\def\tr{{\rm tr}} \def\Tr{{\rm Tr}}
\def\lspace{\ifx\answ\bigans{}\else\qquad\fi}
\def\lbspace{\ifx\answ\bigans{}\else\hskip-.2in\fi} 
\def\boxeqn#1{\vcenter{\vbox{\hrule\hbox{\vrule\kern3pt\vbox{\kern3pt
        \hbox{${\displaystyle #1}$}\kern3pt}\kern3pt\vrule}\hrule}}}
\def\mbox#1#2{\vcenter{\hrule \hbox{\vrule height#2in
                \kern#1in \vrule} \hrule}}  
%
\def\CAG{{\cal A/\cal G}}   
\def\CA{{\cal A}} \def\CC{{\cal C}} \def\CF{{\cal F}} \def\CG{{\cal G}}
\def\CL{{\cal L}} \def\CH{{\cal H}} \def\CI{{\cal I}} \def\CU{{\cal U}}
\def\CB{{\cal B}} \def\CR{{\cal R}} \def\CD{{\cal D}} \def\CT{{\cal T}}
\def\e#1{{\rm e}^{^{\textstyle#1}}}
\def\grad#1{\,\nabla\]_{{#1}}\,}
\def\gradgrad#1#2{\,\nabla\]_{{#1}}\nabla\]_{{#2}}\,}
\def\psibar{\overline\psi}
\def\om#1#2{\omega^{#1}{}_{#2}}
\def\vev#1{\langle #1 \rangle}
\def\lform{\hbox{$\sqcup$}\llap{\hbox{$\sqcap$}}}
\def\darr#1{\raise1.5ex\hbox{$\leftrightarrow$}\mkern-16.5mu #1}
\def\lie{\hbox{\it\$}} 
\def\ha{{1\over2}}
\def\half{{\textstyle{1\over2}}} 
\def\roughly#1{\raise.3ex\hbox{$#1$\kern-.75em\lower1ex\hbox{$\sim$}}}

\voffset=-18pt \hoffset=20pt \def\cl{\centerline} \vsize=56pc
\hsize=35pc \tolerance=1000 \def\ni{\noindent} \baselineskip=23pt
\parskip=3pt plus 2pt minus 1pt \def\i{\item} \def\ii{\itemitem}
\raggedbottom \font\fvlr=cmr10 at 14.4pt \font\fvlbx=cmbx10 at 14.4pt
\font\flr=cmr10 at 12pt\font\flbx=cmbx10 at 12pt\font\flss=cmss10 at 12pt
\font\fbx=cmbx10 at 10pt \font\fr=cmr10 at 10pt\font\fss=cmss10 at 10pt
\font\ftt=cmtt10 at 10pt\font\fit=cmti10 at 10pt\font\flit=cmti10 at 12pt
\fss\def\sqr#1#2{{\vcenter{\hrule height.#2pt \hbox{\vrule width.#2pt
   height#1pt \kern#1pt \vrule width.#2pt} \hrule height.#2pt}}}
\def\sq{{\mathchoice\sqr55\sqr55\sqr{2.1}3\sqr{1.5}3}\hskip 1.5pt}
\def\lrhup#1{\buildrel{\leftharpoonup\hskip -8pt\rightharpoonup}\over #1}
\def\sect#1{\vskip 3pt \ni{\fbx #1} \nobreak \vskip 3pt}
\def\subsect#1{\vskip 1pt \ni{\fbx #1} \nobreak \vskip 1pt}
\def\ent#1#2{{{\baselineskip=13pt \vskip 1pt \i{[#1]}{\fr{#2} \vskip 9pt}}}}
\def\rbar{R \hskip -5pt R} \def\perm#1{\hskip 2pt #1 {\rm - perms}}
\def\footstrut{\baselineskip 12pt} \def\tr{\,{\rm Tr}\,} \hfuzz=75pt
\def\eqnn#1{\eqno{\hbox{(#1)}}}\def\iii{\'{\char'20}}
\def\a{\alpha}\def\b{\beta}\def\g{\gamma}\def\G{\Gamma}\def\d{\delta}
\def\D{\Delta}\def\e{\epsilon}\def\ee{\varepsilon}\def\z{\zeta}
\def\th{\theta}\def\TH{\Theta}\def\tth{\vartheta}\def\k{\kappa}
\def\l{\lambda}\def\L{\Lambda}\def\m{\mu}\def\n{\nu}\def\cs{\xi}
\def\Cs{\Xi}\def\p{\pi}\def\P{\Pi}\def\r{\rho}\def\s{\sigma}\def\S{\Sigma}
\def\t{\tau}\def\y{\upsilon}\def\Y{\upsilon}\def\f{\phi}\def\F{\Phi}
\def\x{\chi}\def\ps{\psi}\def\Ps{\Psi}\def\o{\omega}\def\O{\Omega}
\def\vf{\varphi}\def\pa{\partial}\def\da{\dagger}\def\dda{\ddagger}
\def\slash#1{#1 \!\!\! /\, }\def\dd#1{{\pa\over\pa #1}}\def\du{d^{4}\!u\,}
\def\dudv{d^{4}\!ud^{4}\!v\,}\def\dr{differential renormalization}

\ref\fjl{D.Z. Freedman, K. Johnson and J.I. Latorre, {\it Nucl. Phys.}
{\bf B371}(1992)353.}

\ref\bphz{N.N. Bogoliubov and O.S. Parasiuk, {\it Acta Math.}
{\bf 97}(1957)227\semi
K.Hepp, {\it Comm.Math.Phys.} {\bf 2}(1966)301\semi
W. Zimmermann, in {\it Lectures on Elementary Particles
and Quantum Field Theory} (MIT Press, 1970), eds. S. Deser, M. Grisaru
and H. Pendleton.}

\ref\fjmv{D.Z. Freedman, K. Johnson, R. Mu\~n\'oz-Tapia
and X. Vilas\iii s-Cardona, ``A Cut-Off Procedure and Counterterms for
Differential Regularization", MIT preprint CTP\#2099.}

\ref\haagensen{P.E. Haagensen, {\it Mod. Phys. Lett.} {\bf
A7}(1992)893.}

\ref\fgjr{D.Z. Freedman, G. Grignani, K. Johnson and N. Rius,
``Conformal Symmetry and Differential Regularization of the
3-Gluon Vertex", MIT preprint CTP\#1991, to appear in {\it Annals of
Physics(NY)}.}

\ref\hl{P.E. Haagensen and J.I. Latorre, ``Differential Renormalization
of Massive Quantum Field Theories", Univ. of Barcelona preprint
UB-ECM-PF 92/5, to appear in {\it Phys. Lett.} {\bf B}.}

\ref\ramon{R. Mu\~n\'oz-Tapia, ``A New Approach to Lower Dimensional
Gauge Theories", Univ. of Durham preprint, DTP/92/34.}

\ref\schwinger{J. Schwinger, {\it Particles, Sources and Fields}, vols.
1 and 2, (Addison-Wesley, Reading, MA, 1970, 1973).}

\ref\bpz{A.A. Belavin, A.M. Polyakov and A.B. Zamolodchikov, {\it
Nucl. Phys.} {\bf B241}(1984)333.}

\ref\landau{V.B. Berestetskii, E.M. Lifshitz and L.P. Pitaevskii,
{\it Quantum Electrodynamics} (Pergamon Press, 1982).}

\ref\itzu{C. Itzykson and J.B. Zuber, {\it Quantum Field Theory}
(McGraw-Hill, New York, 1980).}

\ref\jauch{J.M. Jauch and F. Rohrlich, {\it The Theory of Photons and
Electrons} (Addison-Wesley, Cambridge, MA, 1955).}

\ref\akhiezer{A.I. Akhiezer and V.B. Berestetskii, {\it Quantum
Electrodynamics} (John Wiley, New York, 1965).}

\ref\gross{D.J. Gross, ``Methods in Field Theory", in {\it Proceedings
of Les Houches Summer School, 1975} (North Holland, 1976), eds. R.
Balian and J. Zinn-Justin.}

\ref\abbott{L.F. Abbott, {\it Nucl.Phys.} {\bf B185}(1981)189.}

\ref\baker{M. Baker and K. Johnson, {\it Physica} {\bf 96A}(1979)120.}


\ref\abj{S.L. Adler, in {\it Lectures on Elementary Particles
and Quantum Field Theory} (MIT Press, 1970), eds. S. Deser, M. Grisaru
and H. Pendleton\semi
R. Jackiw,  ``Field Theoretic Investigations in Current Algebra'',
in {\it Current Algebra and Anomalies} (World Scientific, 1985), eds.
S.B. Treiman, R. Jackiw, B. Zumino and E. Witten.}

\ref\adlerbardeen{S.L. Adler and W. Bardeen, {\it Phys. Rev.}
{\bf 182}(1969)1517.}

\ref\schreier{E.J. Schreier, {\it Phys. Rev.} {\bf D3}(1971)982.}

\ref\dan{D.Z. Freedman, ``Differential Regularization and
Renormalization: Recent Progress", in {\it Proceedings of the Stony
Brook Conference on Strings and Symmetries, 1991}, MIT preprint
CTP\#2020\semi K. Johnson, private communication.}

\null
\vskip 2pc

\pageno=0
\footline={\null \hfill}

\cl{\bf A COMPREHENSIVE COORDINATE SPACE RENORMALIZATION OF}

\cl{\bf QUANTUM ELECTRODYNAMICS TO 2-LOOP ORDER}

\baselineskip=15pt  \vskip 6pc

\cl{Peter E. Haagensen\footnote{$^1$}{e-mail:
hagensen@ebubecm1.bitnet} and
Jos\'e I.~Latorre\footnote{$^2$}{e-mail:
latorre@ebubecm1.bitnet}}

\vskip 2pc

\cl{\it Departament d'Estructura i Constituents de la Mat\`eria}
\cl{\it Facultat de F\iii sica, Universitat de Barcelona}
\cl{\it Diagonal 647, 08028 Barcelona, Spain}

\vskip 7pc

\cl{\bf ABSTRACT}

We develop a coordinate space renormalization of massless Quantum
Electrodynamics using the powerful method of differential renormalization.
Bare one-loop amplitudes are finite at non-coincident external points,
but do not accept a Fourier transform into momentum space. The method
provides a systematic procedure to obtain one-loop renormalized
amplitudes
with finite Fourier transforms
in strictly four dimensions without the appearance of
integrals or the use of a regulator. Higher loops are solved similarly by
renormalizing from the inner singularities outwards to the global one. We
compute all 1- and 2-loop 1PI diagrams, run renormalization group
equations on them and check Ward identities. The method furthermore
allows us to discern a particular pattern of renormalization under which
certain amplitudes are seen not to contain higher-loop leading
logarithms. We finally present the computation of the chiral triangle
showing that differential renormalization emerges as a natural scheme to
tackle $\gamma_5$ problems.

\vfill

\ni{\bf UB-ECM-PF 92-14\hfill June 1992}

\vskip 1pc  \baselineskip=22pt  \eject

\footline={\hss \folio \hss}

\newsec{Introduction}

All one-loop diagrams in Quantum Electrodynamics (QED) are
products of propagators when written in coordinate space and,
therefore, finite for separate external points. Apart from tadpoles,
this is in fact true for any quantum field theory. This striking
observation has been in the mind of the physics
community since long ago though little practical use of it has been
made. The need for renormalization
comes through the fact that the product of propagators is not a
distribution and, thus, has no Fourier transform. In crude terms,
renormalization is a procedure to extend products of distributions
into distributions.

Differential Renormalization (DR) \fjl\ provides a prescription to
implement
this project which is in a sense minimal. The basic idea is to write an
amplitude in position space and then
express it as derivatives of a distribution less divergent at
coincidence points. The typical computation
of primitively divergent Feynman integrals is substituted for solving
trivial
differential equations. This prescription is complemented with the
instruction of using the derivative in front of the distribution as
acting to its left, which is standard in the theory of distributions.

We claim that \dr\ is minimal because it never changes the value of
a primitively divergent Feynman diagram away from its singularities,
and neither does it modify the dimensionality
of space-time  or introduce a regulator. It is
simple
and addresses straighforwardly the concept of renormalization, rather
than regularization. In spirit, it is close to BPHZ renormalization
\bphz , in that
both prescriptions deliver renormalized amplitudes graph by graph.
In the absence of a renormalized lagrangian, DR
needs a supplementary effort to prove unitarity. In \fjmv ,
this problem has been analyzed for $\l\f^4$-theory and perturbative
unitarity proven to 3-loop order.

A number of theories have been used as a test for \dr . In \haagensen ,
the supersymmetric Wess-Zumino model was studied up to three loops with
ease. Conformal invariance in QCD was exploited in \fgjr\ thanks to the
position space nature of \dr . The method also extends to massive
theories \hl\  and to lower dimensional theories \ramon .

Other coordinate space regularization and renormalization methods have
occasionally been used in the past. From Schwinger \schwinger\ to all
the
recent conformal field theory developments \bpz , the advantages of
postponing
loop integration to higher loops and the possibility of exploiting
conformal invariance have led to numerous applications of coordinate
space methods.

In this paper we present a comprehensive study of the differential
renormalization of massless QED up to two loops. There are obvious
reasons to undertake such a computation.  QED is a physical
theory with gauge symmetry. The consistency of the method passes a
stringent test because of the existence of Ward identities (WIs) and
potential problems hidden in overlapping divergences.
It also faces the problem of the chiral anomaly and the correct
treatment of $\gamma_5$.

Our results are in perfect agreement with other computational schemes
(e.g., \schwinger ,\landau ,\itzu ,\jauch ,\akhiezer )
but are obtained in a simpler form. We devote Sections 1 and 2 to
presenting the renormalized amplitudes up to two loops. Whereas several
standard treatments of QED will perform the subtractions on mass-shell,
here renormalized amplitudes are subtracted at an arbitrary and
unphysical scale $M$, due to the fact that we are dealing with
{\it massless} QED; this massless limit exists as a quantum field theory
and provides substantial information about massive QED.
We pay special
attention to the way \dr\ uses the basic WIs of
the theory and how they are repeatedly checked in two-loop
computations. Deeply rooted in the nature of DR lies the idea that
each Feynman diagram has as many hidden scales as singular regions.
Some of these scales are fixed by
WIs. The rest are renormalization-scheme prescriptions. Keeping
all this freedom manifest allows the
method to ``predict" that the renormalization group
equations will display a particular organization. This
``structured renormalization group", discussed in Section 3, provides
a deep insight into
the absence of promotion of leading logs in some diagrams, which is
seen as accidental in standard treatments.
Finally, we review the calculation of the chiral anomaly in
our regulator-free way in Section 5. We also compute the
absence of infinite renormalizations (i.e., absence of $\ln$'s) at two
loops of the triangle
amplitude. As differential renormalization never leaves four dimensions,
it stands as a good candidate to work with $\gamma_5$-related
observables. The anomaly is just one instance.

\newsec{One-loop Renormalization}
\subsect{1.1 Conventions and definitions}

In $d=4$ Euclidean space, massless QED is described by the Lagrangian:
\eqn\lagr{ {\cal L}={1\over 4}F_{\m\n}^2 + {1\over 2a}(\pa_\m A_\m)^2 +
\psibar(\pa\!\!\! / + ieA\!\!\! / )\ps , }
where $A_\m (x)$ and $\ps (x)$ are the usual $U(1)$ gauge and
Fermi fields, $F_{\m\n}(x)=\pa_\m A_\n (x)-\pa_\n A_\m (x)$ is the gauge
field strength, $a$ is a gauge fixing parameter, and $\g$-matrices
satisfy the usual Clifford algebra, $\{\g_\m ,\g_\n\} =2\d_{\m\n}$.

Loop diagrams are calculated with the following Feynman rules:

\ni{\it i)} the gauge field and massless fermion propagators
are, respectively,

\vskip 28pt

\vbox{\baselineskip=35pt plus 2pt minus 1pt
$$\eqalign{
G_{\m\n}^{(0)}(x-y;a)&={1\over 4\p^2}\left[\left({1+a\over 2}\right)
{\d_{\m\n}\over (x-y)^2}+(1-a){(x-y)_\m (x-y)_\n\over (x-y)^4}\right]\cr
S^{(0)}(x-y)&=-{1\over 4\p^2}\g_\m \pa_\m {1\over (x-y)^2},}$$  }

\ni{\it ii)} to each vertex is associated a factor $ie\g_\m$,

\ni{\it iii)} internal coordinates are integrated over,

\ni{\it iv)} closed fermion
loops are multiplied by a factor $-1$, and

\ni{\it v)} diagrams have no symmetry factors.

\ni We also use the convention that for derivative operators ${\cal O}=
\pa_\m , \pa\!\!\! /,\sq ,$ etc., we take ${\cal O}f(x-y)$ to mean ${\cal
O}(x)f(x-y)$, unless explicitly stated otherwise.

Renormalization of QED
entails the renormalization of only three different vertex functions:
the 2-point vacuum polarization, $\P_{\m\n}(x-y)$, the 2-point fermion
self-energy, $\S (x-y)$, and the 3-point vertex, $V_\m(x-z,y-z)$. These
will contribute to the effective action in the following way:
$$ \G_{\rm eff}=\int d^4\!x d^4\!y \left\{ {1\over 2}A_\m (x)
\left[\left( (1-{1\over a})\pa_\m\pa_\n -\d_{\m\n}\sq\right)\d^{(4)}(x-y)
-\P_{\m\n}(x-y)\right]A_\n (y)\right.$$
$$+\left.\psibar (x)\left(\slash{\pa}
\d^{(4)}(x-y)-\S (x-y)\right)\ps (y)\right\}$$
\eqn\sigue{+\int d^4\!x d^4\!y d^4\!z\, \psibar (x)\left[ie\g_\m
\d^{(4)}(x-y)
\d^{(4)}(x-z)+V_\m(x-z,y-z)\right]\ps (y)A_\m(z)+\cdots .}

\vskip 9pt

As defined above, these renormalization parts will contain only 1PI loop
contributions and $\P_{\m\n}$ and $\S$ will, furthermore, lead to the
following full photon and electron propagators:
\eqn\propag{\eqalign{
G_{\m\n}&=G_{\m\n}^{(0)}+G_{\m\r}^{(0)}\cdot\P_{\r\s}\cdot G_{\s\n}^{(0)}
+G_{\m\a}^{(0)}\cdot\P_{\a\b}\cdot G_{\b\r}^{(0)}\cdot\P_{\r\s}\cdot
G_{\s\n}^{(0)}+\cdots =(G_{\m\n}^{(0)-1}-\P_{\m\n})^{-1}\cr
S&=S^{(0)}+S^{(0)}\cdot\S\cdot S^{(0)}+S^{(0)}\cdot\S\cdot S^{(0)}\cdot
\S \cdot S^{(0)}+\cdots =(S^{(0)-1}-\S )^{-1},}  }
where the dot indicates a convolution.
To all loop orders, renormalized vertex parts satisfy renormalization
group (RG) equations \gross\ and, as a consequence of gauge invariance,
the following Ward identities:
\eqn\wia{ \pa_{\m} \P_{\m\n}(x-y)=0,}
\eqn\wib{
{\pa\over \pa z^\m}V_\m(x-z,y-z)=-ie[\d^{(4)}(z-x)-\d^{(4)}(z-y)]\S
(x-y).}
The specific
coefficients in the second Ward identity are uniquely fixed by the
tree-level
identity solved by $V_\m^{(0)}=ie\g_\m\d^{(4)}(x-y)\d^{(4)}(x-z)$ and
$S^{(0)-1}=\slash{\pa}\d^{(4)}(x-y)$.

In what follows, we shall find the 1- and 2-loop corrections to
$\P_{\m\n}$, $\S$ and $V_\m$. We will verify the above Ward identities,
and the requirement that renormalization parts satisfy
renormalization group equations will allow us to calculate the
renormalization group functions of QED (beta-functions and anomalous
dimensions).

\vskip 1pc

\subsect{1.2 Renormalization}
The bare 1-loop vacuum polarization (Fig. 1a) reads:
\eqn\barevacpol{\eqalign{
\P_{\m\n}^{(1){\rm bare}}(x-y)&=
-\left({ie\over 4\p^2}\right)^2\tr
\left(\g_\m\slash{\pa}{1\over(x-y)^2} \g_\n
\slash{\pa}{1\over (y-x)^2}\right)\cr &
=-{\a\over 3\p^3}(\pa_\m\pa_\n -\d_{\m\n}\sq )\,{1\over (x-y)^4},} }
with $\a=e^2/4\p$ the fine structure constant. Using the basic DR
identity (cf. Appendix A for all the identities needed in renormalizing
amplitudes up to two loops):
\eqn\boxlog{ {1\over x^4}=-{1\over 4}\ \sq\ {\ln x^2M^2\over x^2}\ ,}
we find the renormalized value of the one-loop vacuum polarization:
\eqn\oneloopvacpol{\P_{\m\n}^{(1)}(x-y)={\a\over 12\p^3}(\pa_\m\pa_\n
-\d_{\m\n}\sq )\, \sq {\ln (x-y)^2M_{\P}^2\over (x-y)^2}\ ,}
with the proviso that the total derivatives above, as well as
in all other renormalized amplitudes throughout, are to be understood
as acting to the left.
Here and below we adopt the convention of appending a subscript
($\P$,
$\S$ or $V$) to the renormalization scales appearing in the different
renormalization parts, since these scales are {\it a priori} independent
(but eventually related through Ward identities, as we shall see). This
quantity gives us then the renormalized $AA$ 2-point function:
\eqn\gaa{
\G_{\m\n}^{AA}(x-y)=
\left( (1-{1\over a})\pa_\m\pa_\n -\d_{\m\n}\sq\right)\d^{(4)}(x-y)
-\P_{\m\n}^{(1)}(x-y).
}
We now impose that it satisfy the usual RG equation:
\eqn\rgvacpol{
\left(M_\P{\pa\over \pa M_\P}+\b (\a){\pa\over \pa\a }+\g_a(\a){\pa\over
\pa a}-2\g_A(\a)\right)\G_{\m\n}^{AA}(x-y)=0,
}
where $\b (\a)$ is the QED $\b$-function, $\g_A(\a)$ is the
anomalous
dimension of the gauge field, and $\g_a(\a)$ a ``$\b$-function''
associated to the running of the gauge parameter $a$. This is then
easily seen to lead to the 1-loop values:
\eqn\anoa{\g_A(\a)={\a\over 3\p}$$ and $$\g_a(\a)=-{2a\a\over 3\p}}
($\b (\a)$ has dropped out above because it is of higher order in $\a$).

It is well-known that the above computation can be used in the framework
of the background field method to find the beta-function of QED
\abbott . It turns out that the relation $\b (\a )=2\a
\g (\a )$ holds to all orders. Therefore, we can anticipate that
the beta function at one-loop is
\eqn\bfmbeta{\b (\a )={2\a^2\over 3\p }\, .}
This result will later be confirmed by an independent computation of the
one-loop vertex.

The bare 1-loop fermion self-energy (Fig. 1b) is:
\eqn\bareselfenergy{\S^{(1){\rm bare}}(x-y)=\left({e\over
4\p^2}\right)^2 \g_\m\ \slash{\pa}{1\over(x-y)^2}\ \g_\n
\left[{1+a\over 2}
{\d_{\m\n}\over (x-y)^2}+(1-a){(x-y)_\m (x-y)_\n\over (x-y)^4}\right].
}
The renormalization of this amplitude is straightforward, apart from
the following important subtlety: if we use the same mass scale $M_\S$
in renormalizing the two parts coming from the two different pieces in
the photon propagator, we will end up with a renormalized amplitude
which is directly proportional to the gauge parameter $a$; now, in
Landau gauge ($a=0$), it would then vanish entirely, and it is easy to
see
this would be inconsistent with the Ward identity. It is in fact a known
property of QED that, in Landau gauge, although the bare 1-loop fermion
self-energy vanishes, the renormalized self-energy does not. In momentum
space, it is equal to a finite constant times $\slash{q}$, and this is a
reflection of the ambiguity inherent in the linearly divergent integral
defining the amplitude. This turns out to be transparent in differential
renormalization: in renormalizing the two pieces in question, no one
tells us we should take the same mass scale $M_\S$, and so we do not. We
renormalize the first piece (the Feynman gauge part) with a scale
$M_\S$, and the second piece with a scale $M'_\S$ related to the first
one by $\ln M'_\S=\ln M_\S -\l$. This then leads to the following
renormalized 1-loop fermion self-energy:
\eqn\oneloopselfenergy{\eqalign{\S^{(1)}(x-y)=&{\a\over
16\p^3}\slash{\pa}\,\sq\left(
{a\ln (x-y)^2M_{\S}^2+\l (1-a)/2\over (x-y)^2}\right)\cr
=&{a\a\over 16\p^3}\slash{\pa}\,\sq {\ln (x-y)^2M_{\S}^2\over
(x-y)^2}-{\l\a (1-a)\over 8\p}\slash{\pa}\d^{(4)}(x-y)\, .}}
As expected, in Landau gauge, all that survives is precisely the
contact term mentioned above. We will see shortly that the Ward identity
will not only determine
a relation between the self-energy and vertex mass scales, but also a
unique value for $\l$.
The renormalized $\psibar\ps$ 2-point function,
\eqn\psibarpsvertex{\G^{\psibar\ps}(x-y)
=\slash{\pa}\d^{(4)}(x-y)-\S^{(1)}(x-y),}
then satisfies the RG equation:
\eqn\psipsirg{
\left(M_\S{\pa\over \pa M_\S}+\b (\a){\pa\over \pa\a }+\g_a(\a){\pa\over
\pa a}-2\g_\ps(\a)\right)\G^{\psibar\ps}(x-y)=0,
}
and this will lead to the following 1-loop fermion anomalous
dimension:
\eqn\psanomdim{\g_\ps(\a)={a\a\over 4\p}}
(here, both $\b$ and $\g_a$ drop out).

Finally, we now renormalize the 1-loop vertex (Fig. 1c). Its bare
expression is:
\eqn\oneloopvertex{\eqalign{
&V_\m^{(1){\rm bare}}(x-z,y-z)=\cr
&=\left({ie\over 4\p^2}\right)^3
\left(\g_\r\g_a\g_\m\g_b\g_\s\right)\pa_a{1\over (x-z)^2}
\pa_b{1\over (z-y)^2}
\left[\left({1+a\over 2}\right)
{\d_{\r\s}\over (x-y)^2}+(1-a){(x-y)_\r (x-y)_\s\over (x-y)^4}\right]\cr
&={-i\over 32\p^3}\left({\a\over \p}\right)^{3/2}
\left(\g_\r\g_a\g_\m\g_b\g_\s\right)\pa_a{1\over (x-z)^2}
\pa_b{1\over (z-y)^2}
\left[(a-1)\pa_\r\pa_\s +\d_{\r\s}\sq \right]\ln (x-y)^2\m^2.
}}

Like for the fermion self-energy, the photon propagator in a
generic gauge leads to two pieces to be renormalized, in principle with
mass scales $M_V$ and $M'_V$. However, it turns out that this need not
be done here, that is, the choice of the same $M_V$
throughout will not lead to any inconsistencies (still, one may choose
different scales at will; this would simply correspond to
different choices of scheme).
We briefly sketch the procedure involved in renormalizing the expression
above. In Feynman gauge ($a=1$) it is most transparent \fjl : there are
three propagators $1/(x-x')^2$ forming a triangle, with derivatives
acting on two of them. This has dimension $L^{-8}$ and therefore it is
log divergent in the ultraviolet. One first integrates by parts until
the two derivatives are acting on the same leg, say, $\pa_a\pa_b
{1\over (x-y)^2}$, and then subtracts and adds a trace piece thus:
$(\pa_a\pa_b -{1\over 4}\d_{ab}\sq ){1\over (x-y)^2}+{1\over
4}\d_{ab}\sq {1\over (x-y)^2}$. The surface terms, with total
derivatives
acting {\it outside}, is power counting $L^{-7}$ and thus finite, and
the traceless combination of derivatives is also finite (due to
tracelessness). The divergence has been isolated in the term
${1\over 4}\d_{ab}\sq {1\over (x-y)^2}=-\p^2\d_{ab}\d^{(4)}(x-y)$,
which turns
out to be easily renormalizable through the DR identity, Eq.\boxlog .
For
a generic gauge, the same principle of separating divergent terms into
trace and traceless pieces applies, only in this case the $\g$-matrix
structure in front complicates things a little: we integrate $\pa_a$ and
$\pa_b$ above by parts onto the photon leg, and then look for the
coefficient $A$ to make the expression
\eqn\someexpression{
\left(\g_\r\g_a\g_\m\g_b\g_\s\right)(\pa_a\pa_b +A\d_{ab}\sq )
\left[(a-1)\pa_\r\pa_\s +\d_{\r\s}\sq \right]
}
a traceless (and thus finite) combination of derivatives. The
appropriate value is $A=-a/(a+3)$; adding and subtracting that trace
piece, we are able to find the renormalized expression for the 1-loop
vertex:
\eqn\renvertex{\eqalign{
&V_\m^{(1)}(x-z,y-z)=\cr
&={-i\over 32\p^3}\left({\a\over \p}\right)^{3/2}
\left\{\left(\g_\r\g_a\g_\m\g_b\g_\s\right)
{\pa\over\pa x^a}\left[{1\over (x-z)^2}\pa_b{1\over (z-y)^2}
\left[(a-1)\pa_\r\pa_\s +\d_{\r\s}\sq\right]\ln (x-y)^2\m^2
\right]\right.\cr
&+4\left[(a-1)\g_\m\g_b -2\g_b\g_\m\right]{\pa\over\pa y^b}
\left[{1\over (x-z)^2(z-y)^2}\slash{\pa}{1\over (x-y)^2}\right]\cr
&\left. -{16\over (x-z)^2(z-y)^2}\g_\s\left[\pa_\m\pa_\s
-{1\over 4}\d_{\m\s}\sq\right]{1\over (x-y)^2}+4\p^2a\g_\m\sq
{\ln (x-z)^2M_{V}^2\over (x-z)^2}\d^{(4)}(x-y)\right\}.}}
We remark here on the fact that the renormalization piece above
(containing $\ln M_{V}^2$) is directly proportional to the gauge
parameter $a$, and this is also true of $\S^{(1)}(x-y)$ found above. This is
a reflection of the well-known fact that, apart from vacuum polarization
infinities, QED is 1-loop finite in Landau gauge.

The renormalized 3-point vertex
\eqn\rentp{\G_\m^{\psibar
A\ps}(x-z,y-z)=ie\g_\m\d^{(4)}(x-y)\d^{(4)}(x-z)+ V_\m^{(1)}(x-z,y-z)}
satisfies the RG equation:
\eqn\rgtp{
\left(M_V{\pa\over \pa M_V}+\b (\a){\pa\over \pa\a}+\g_a (\a){\pa\over
\pa a} -2\g_\ps (\a) -\g_A(\a)\right)\G_\m^{\psibar A\ps}(x-z,y-z)=0}
with values found previously for $\g_\ps$ and $\g_A$, and
\eqn\prvalues{ \b (\a )={2\a^2\over 3\p}\, .}
(and, again, $\g_a$ is of higher order than we are considering).
This confirms the result from the background field method shown
previously. Naturally, in the above equation
the $a$-dependent pieces ($M{\pa\over\pa M}$ and $\g_\ps$) and
the $a$-independent pieces ($\b{\pa\over\pa\a}$ and $\g_A$) cancel
separately.

We now consider the Ward identity for $V_\m$ and $\S$.

\vskip 1pc

\subsect{1.3 Ward Identity}

For separate points $x\ne y\ne z$, Eq.\wib\ is trivially verified.
The subtlety involved in a coordinate space Ward identity is, however,
its validity at contact. In order to verify Eq.\wib , then, we must be
careful in particular not to lose any contact terms due to formal
manipulations with delta functions, and the best way to guarantee a
correct procedure is to integrate it over one of the external
variables. This was done in \fjl\ for Feynman gauge, and we use the same
procedure here for an arbitrary gauge. The integrated form of the Ward
identity is:
\eqn\intwi{\int d^4\!y{\pa\over\pa z^\m}V^{(1)}_\m(x-z,y-z)=ie\S
(x-z).}
Integrating the expression for the renormalized vertex, Eq.\renvertex ,
over $y$, one finds:
\eqn\wiresult{\int d^4\!y V^{(1)}_\m(x-z,y-z)=
{-i\over 4\p}\left({\a\over \p}\right)^{3/2}
\left[\left(\pa_\m\slash{\pa}-{1+2a\over 4}\g_\m\sq\right)
{1\over (x-z)^2}+{a\over 2}\g_\m\sq
{\ln (x-z)^2M_{V}^2\over (x-z)^2}\right].}

Acting now with the $z$ derivative:
\eqn\zder{\int d^4\!y \dd{z^\m}V^{(1)}_\m(x-z,y-z)={ie^3\over 64\p^4}
\slash{\pa}\sq\left({a\ln (x-z)^2M_{V}^2+(3/2-a)\over (x-z)^2}\right)
\, .}
Now, we compare this with:
\eqn\morewi{ie\S^{(1)}(x-z)={ie^3\over 64\p^4}\slash{\pa}\,\sq\left(
{a\ln (x-z)^2M_{\S}^2+\l (1-a)/2\over (x-z)^2}\right)\, .}
Setting $a=0$ gives $\l =3$, and then, for any $a$, we also find:
\eqn\mrelation{\ln{M_{\S}^2\over M_{V}^2}={1\over 2}\, . }

We see then that a Ward identity relating two renormalized amplitudes
will, in the context of differential renormalization, enforce a relation
between the scales that renormalize the amplitudes.
It is an important point we will analyze further that the
converse is also true, namely, mass scales that renormalize amplitudes
{\it not} related by a symmetry are not related, and this in turn also
enforces constraints in the renormalization of these amplitudes. At two
loops, in particular for the anomalous triangle, the use of the above
mass relation will be crucial for the consistency of our calculations.

\newsec{Two-loop Renormalization}

At two loops, the 1PI diagrams renormalizing $\P_{\m\n}$, $\S$ and
$V_\m$ are depicted in Fig. 2. Throughout this
section, we will perform all calculations in Feynman gauge ($a=1$);
since $\g_a{\pa\over \pa a}$ acting on any 2-loop diagram will be
one order higher in $\a$, this will not affect the verification of RG
equations.

\vskip 1pc

\subsect{2.1 Vacuum Polarization}

We begin with the simplest of the two vacuum polarization diagrams
(Fig. 2a), and from now on we will use translation invariance to set one
external point to zero in all diagrams, for the sake of simplicity. Its
bare expression is:
\eqn\twoloopvac{ \P_{\m\n}^{(2a){\rm bare}}(x)={(ie)^2\over
(4\p^2)^3}\int\dudv
\tr \left( \g_\m\slash{\pa}{1\over (x-u)^2}\S^{(1)}(u-v)
\slash{\pa}{1\over v^2}\g_\n\slash{\pa}{1\over x^2}\right).
}
Standard manipulations lead to:
\eqn\vacpoltwofinal{\P_{\m\n}^{(2a){\rm bare}}(x)={e^4\over 48\p^6}
\left[{1\over 4}(\pa_\m\pa_\n -\d_{\m\n}\sq)\left({\ln x^2M_\S^2-1/6
\over x^4}\right)-{\d_{\m\n}\over x^6}\right].}
We
note that this diagram is not transverse by itself; the non-transverse
piece will be cancelled when we consider the entire 2-loop vacuum
polarization. The renormalization now proceeds with the straightforward
use of the DR identities listed in Appendix A, and we find:

\eqn\pmn{
\P_{\m\n}^{(2a)}(x)=-{1\over 96\p^2}{\left(\a\over\p\right)}^2
\left\{(\pa_\m\pa_\n -\d_{\m\n}\sq)\sq\left({\ln^2x^2M_\S^2+{5\over 3}
\ln x^2M^2\over x^2}\right)
-\d_{\m\n}\sq\sq{\ln x^2M^2\over x^2}\right\},}
where $M$ is a new renormalization mass parameter appearing at two
loops. The diagram with the fermion self-energy inserted
on the lower leg of the loop will have the same value as this
one. We note here that the $\ln M_\S^2$ coming from the 1-loop
self-energy subdivergence
has been promoted to a $\ln^2M_\S^2$ at two loops: this is the
self-consistency of the renormalization group at work.
We will choose everywhere the same 2-loop mass scale $M$.
In renormalization terms, this simply corresponds to some choice of
renormalization scheme: because these appear as $\sq{\ln x^2M^2\over
x^2}$, different
values of $M$ will lead to amplitudes that differ by finite contact
terms or, in other words, by finite renormalizations (and this will be
true for the new $M$'s appearing at each loop order). Of course, once
we set the (2-loop) $M$'s we like for the renormalization of $\S$
and $V_\m$ in particular, they will subsequently be related by the Ward
identity (but this will not be of concern to us here, as we will not
compute the 2-loop mass relation). Throughout this section, then, we
will only care to distinguish
between $M_\S$, $M_V$ or $M_\P$ and other mass scales $M$ when the former
appear as promoted $\ln$'s, i.e., as $\ln^2M_\S$, etc..

The other vacuum polarization diagram, Fig. 2b, is the most difficult
integral we have had to perform.
The bare expression is:
\eqn\pmntb{ \P_{\m\n}^{(2b){\rm bare}}(x-y)={(ie)^4\over
(4\p^2)^5}\int\dudv
\tr \left( \g_\m\slash{\pa}{1\over (x-u)^2}\slash{\pa}{1\over
(u-y)^2}\g_\n \slash{\pa}{1\over (v-y)^2}\slash{\pa}{1\over
(v-x)^2}\right) {1\over (u-v)^2}.}

There are potential problems related to overlapping
divergences in this amplitude, and so we must examine how differential
renormalization
deals with them. There are two subdiagrams which contain log
singularities related to the regions $u\sim v\sim x$ and $u\sim v\sim
0$. We remove both divergences by pulling out derivatives in $x$ and $y$
in a symmetric way as we did in the case of the one-loop
vertex, Eq.\renvertex . More explicitly, the important intermediate step
is \eqn\intermediatestep{\eqalign{ &
\partial_d{1\over(x-u)^2}\partial_a{1\over (y-u)^2}
\partial_b{1\over (y-v)^2}\partial_c{1\over (x-v)^2}{1\over (u-v)^2}
=\cr
[ {\partial\over\partial x^d}   (  &{1\over
(x-u)^2}\partial_c{1\over (x-v)^2}    ) -{1\over (x-u)^2}(
\partial_d\partial_c-{\d_{cd}\over 4}\sq){1\over (x-v)^2}+
\pi^2\d_{cd} {\delta^4(x-v)\over (x-u)^2}] {1\over (u-v)^2}\cr
[ {\partial\over\partial y^b}( &{1\over
(y-v)^2}\partial_a{1\over (y-u)^2})-{1\over (y-v)^2}(
\partial_b\partial_a-{\d_{ba}\over 4}\sq){1\over (y-u)^2}+
\pi^2\d_{ab} {\delta^4(y-u)\over (y-v)^2}]\,\, . }}
This manipulation makes it evident that each subdivergence is cured
separately thanks to the fact that in coordinate space the external
points are kept apart (whereas, in momentum space, the momentum
integral
would make the two singularities overlap). The rest of the computation,
carried out with the help of Feynman parameters, although somewhat
lengthy,
is conceptually simple since only a global divergence needs further
correction. The final
renormalized value is:
\eqn\pmntbx{
\P_{\m\n}^{(2b)}(x)=-{1\over 48\p^2}{\left(\a\over\p\right)}^2
\left\{-(\pa_\m\pa_\n -\d_{\m\n}\sq)\sq\left({\ln^2x^2M_V^2+{17\over 3}
\ln x^2M^2\over x^2}\right)
+\d_{\m\n}\sq\sq{\ln x^2M^2\over x^2}\right\},}

Taking into account the mass relation (Ward identity), Eq.\mrelation ,
the entire 2-loop renormalized vacuum polarization
reads:
\eqn\morepmntb{
\P_{\m\n}^{(2)}(x)=2\P_{\m\n}^{(2a)}(x)+\P_{\m\n}^{(2b)}(x)=
{1\over 16\p^2}{\left(\a\over\p\right)}^2
(\pa_\m\pa_\n -\d_{\m\n}\sq)\sq{\ln x^2M^2\over x^2}\, .
}
This is automatically transverse, and furthermore the $\ln^2$
contributions have cancelled. This is an important point which we will
further elaborate in Section 3.

The 2-loop renormalized $AA$ 2-point function
\eqn\aavertex{
\G_{\m\n}^{AA}(x)=
\left( (1-{1\over a})\pa_\m\pa_\n -\d_{\m\n}\sq\right)\d^{(4)}(x)
-\P_{\m\n}^{(1)}(x)-\P_{\m\n}^{(2)}(x)}
will satisfy an RG equation, Eq.\rgvacpol , with $M_\P\dd{M_\P}$
substituted for $M_\P\dd{M_\P}+M\dd{M}$ (in general, an
RG equation will include the sum of the derivatives
w.r.t. all masses present in an amplitude). This will yield the 2-loop
RG functions:
\eqn\gaatwoloops{\g_A(\a)={\a\over 3\p}+{\a^2\over 4\p^2}$$
and $$\g_a(\a)=-{2a\a\over 3\p}-{a\a^2\over 2\p^2}=-2a\g_A(\a).}
As mentioned before, the same calculation in the background field method
would have led us to the 2-loop beta function, which we will derive and
present independently through the calculation of the 2-loop vertex in
Sec. 2.3.

\vskip 1pc

\subsect{2.2 Fermion Self-Energy}

Next, we consider 2-loop contributions to the fermion self-energy, shown
in Figs. 2c-2e. In order to eventually verify RG equations and Ward
identities, we also give the logarithmic mass derivative $M\dd{M}$
acting on each one of the amplitudes. The first diagram is that of
Fig. 2c:
\eqn\stc{\S^{(2c){\rm bare}}(x)=-{(ie)^2\over (4\p^2)^3}\int\dudv
{1\over x^2}\,\g_\m\,\slash{\pa}{1\over (x-u)^2}\,\S^{(1)}(u-v)\,
\slash{\pa}{1\over v^2}\,\g_\m \,\, .
}

Again, integration by parts, properties of $\g$-matrices and DR
identities lead to the renormalized value:
\eqn\stcx{\S^{(2c)}(x)=
-{1\over 128\p^2}{\left(\a\over\p\right)}^2\slash{\pa}
\sq\left({\ln^2x^2M_\S^2+\ln x^2M^2\over x^2}\right),}
where we have also used the identity
\eqn\idneti{{1\over x^2}\pa_a\left({\ln x^2M_\S^2\over x^2}\right)=
{1\over 2}\pa_a\left({\ln x^2M_\S^2-1/2\over x^4}\right).}
The mass derivative of this amplitude is:
\eqn\mdmstc{ M\dd{M}\,\,\S^{(2c)}=-{\a\over 2\p}\,\,\S^{(1)}-
{\a^2\over 16\p^2}\,\,\S^{(0)}.}
The next contribution is that of Fig.  2d:
\eqn\std{\S^{(2d){\rm bare}}(x)=-{(ie)^4\over (4\p^2)^5}\int\dudv
{1\over u^2}{1\over (x-v)^2}\g_\m\slash{\pa}{1\over(x-u)^2}
\g_\n\slash{\pa}{1\over(u-v)^2}\g_\m\slash{\pa}{1\over v^2}\g_\n \,\, .
}
This amplitude is solved by using vertex-like manipulations. It turns
out that no new integrals are needed apart from those appearing
in the two-loop
vacuum polarization diagram 2b. We also make use of the
following trick
$$\sq\left({1\over (v-x)^2}{1\over v^2}\right)=-4\p^2\left(
{\d^{(4)}(v-x)\over v^2}+{\d^{(4)}(v)\over (v-x)^2}\right)+
2\pa_a{1\over (v-x)^2}\pa_a{1\over v^2},$$
which simplifies part of the computation.
The renormalized value we finally get is:
\eqn\stdx{\S^{(2d)}(x)={1\over
64\p^2}\left({\a\over\p}\right)^2\slash{\pa}
\sq {\ln^2x^2M_V^2\over x^2},}
and the mass derivative of this amplitude is:
\eqn\mdmstd{ M\dd{M}\,\, \S^{(2d)}={\a\over \p}\,\,\S^{(1)}-
{\a^2\over 8\p^2}\,\,\S^{(0)},}
where we have used the 1-loop mass relation to express $\S^{(1)}$ with
the mass scale $M_\S$ rather than $M_V$.

We finally present the last of the 2-loop fermion self-energy diagrams
(Fig. 2e):
\eqn\ste{\S^{(2e){\rm bare}}(x)=-{(ie)^2\over (4\p^2)^3}\,\g_\m\,
\slash{\pa}
{1\over x^2}\,\g_\n\int\dudv{1\over
(x-v)^2}\,\P^{(1)}_{\m\n}(v-u)\,{1\over u^2}\,\, . }
The renormalization of this amplitude is straightforward, and
we only point out that while there is in principle the possibility
that the $\ln M_\P$ in $\P^{(1)}_{\m\n}$ would get promoted to a
$\ln^2M_\P$, the $\g$-matrices in the amplitude, together with the
transverse operator in $\P^{(1)}_{\m\n}$, conspire to simply cancel all
$\ln^2$'s. This is again an instance of the feature we have seen
previously in $\P^{(2)}_{\m\n}$, namely, the apparently unexpected
cancellation of some particular divergences.
The final, renormalized result is:
\eqn\stex{\S^{(2e)}(x)=
-{1\over 32\p^2}{\left(\a\over\p\right)}^2\slash{\pa}
\sq{\ln x^2M^2\over x^2}\, .}
The mass derivative of this amplitude will be:
\eqn\mdmste{ M\dd{M}\,\,\S^{(2e)}=-{\a^2\over 4\p^2}\,\,\S^{(0)}.}
Whereas we would generally expect the mass derivative of a 2-loop
amplitude to generate 1-loop and tree-level amplitudes, we see this does
not happen here. In Section 3, we will examine this more closely.

We now add all these amplitudes to find the 2-loop renormalized
$\psibar\ps$ 2-point function:
\eqn\rentwolooppsipsi{\eqalign{
\G^{\psibar\ps}(x)=&\slash{\pa}\d^{(4)}(x)-\S^{(1)}(x)
-\S^{(2c+2d+2e)}(x)\cr
=&\slash{\pa}\d^{(4)}(x)
-{a\over 16\p^2}\left({\a\over\p}\right)
\slash{\pa}\,\sq{\ln x^2M_{\S}^2\over x^2}
-{1\over 128\p^2}\left({\a\over\p}\right)^2
\slash{\pa}\,\sq\left({\ln^2x^2M_{\S}^2-7\ln x^2M^2\over
x^2}\right)\, .} }
 We have the 1-loop correction in a generic gauge $a$, and the 2-loop
terms in Feynman gauge. This is written thus because in RG equations we
will need $\dd{a}$ on the 1-loop term but not on the 2-loop terms. The
2-loop fermion anomalous dimension given by the RG equations is (in
Feynman gauge):
\eqn\gpsa{\g_\ps(\a)={\a\over 4\p}-{3\a^2\over 32\p^2}\, .}

\vskip 1pc

\subsect{2.3 Vertex}

We now turn to the computation of 2-loop corrections to the gauge
coupling vertex, shown in Figs. 2f-l. Many of the integrals are
extremely difficult to perform analytically, but fortunately
they need not be done for the purposes of verifying RG equations and
Ward identities (although we will not present the 2-loop relation
between $M_V$ and $M_\S$). The parts of these diagrams that do need to
be computed fully are the renormalization pieces from both global and
internal divergences, and those are reasonably simple to calculate.
Because the expressions for the renormalized amplitudes are
somewhat lengthy, we present them in Appendix B. Here, instead, we
will explain briefly the different techniques we have used and subtleties
involved as we go along. Again, for the purpose of later verifying RG
equations
and Ward identities, we also present here the result of the logarithmic
mass derivative $M\dd{M}$ acting on each one of the renormalized
amplitudes. We set $z=0$ in all amplitudes.

The first diagram we consider is Fig. 2f:
\eqn\vtf{V_\m^{(2f){\rm bare}}(x,y)=-{(ie)^3\over (4\p^2)^4}{1\over
(x-y)^2}
\int\dudv\g_\r\,\slash{\pa}{1\over x^2}\,\g_\m\,\slash{\pa}{1\over u^2}
\S^{(1)}(u-v)\slash{\pa}{1\over (v-y)^2}\,\g_\r\, .}

The standard integrations by parts and separation into trace and
traceless pieces which were used to renormalize the 1-loop vertex are
used here as well, without any added complication. We also need
to consider the diagram identical to the one above, but with the
self-energy
insertion on the other fermion leg, Fig. 2g. The procedure is identical
to the one used above, and its contribution to the RG equations at two
loops is also the same. From the renormalized expression given in
Appendix B, we can calculate the mass derivative of this amplitude:
\eqn\mdmvtf{ M\dd{M}\,\, [V_\m^{(2f)}+V_\m^{(2g)}]=-{\a\over\p}\,\,
V_\m^{(1)}- {3\a^2\over 8\p^2}\,\, V_\m^{(0)}.}
Here and below, we are always taking $V_\m^{(1)}$ in Feynman gauge,
unless otherwise explicitly stated;
we must also be careful to express the result in terms of $M_V$ and not
$M_\S$ (in the above, for instance, neglecting this would lead to a
different coefficient for the contact piece $V_\m^{(0)}$).

The second vertex diagram we compute, Fig. 2h, contains a vacuum
polarization insertion:
\eqn\vth{V_\m^{(2h){\rm bare}}(x,y)={(ie)^3\over(4\p^2)^4}
\left(\g_\r\g_a\g_\m\g_b\g_\s\right) \pa_a{1\over x^2}
\pa_b{1\over y^2}\int\dudv {1\over (x-u)^2}
\P_{\r\s}^{(1)}(u-v) {1\over (v-y)^2}\, . }
 To renormalize this, we integrate the derivatives $\pa_a$ and
$\pa_b$ by parts onto the photon leg. The transversality of
$\P_{\m\n}^{(1)}$ and the gamma structure in front will arrange things
so that the resulting combination of derivatives will automatically be
traceless, thus obviating the need for further renormalization and
(again) avoiding the promotion of $\ln M_\P$. This in turn leads to a
peculiar form for the mass derivative:
\eqn\mdmth{ M\dd{M}\,\, V_\m^{(2h)}=-{2\a\over 3\p}\,\,
V_\m^{(1)}(a=0).}

Now, we are left with the more difficult diagrams, which cannot be
computed in closed form. Starting with the diagram of
Fig. 2i, its bare value is:
\eqn\vti{\eqalign{ V_\m^{(2i){\rm bare}}(x,y)=&
{-(ie)^5\over(4\p^2)^6}\left(\g_\r\g_a\g_\s\g_b\g_\m\g_c\g_\s\g_d\g_\r
\right)\times\cr &{1\over (x-y)^2}
\int\dudv {1\over (u-v)^2}\pa_a{1\over (x-u)^2} \pa_b{1\over u^2}
\pa_c{1\over v^2}\pa_d{1\over (v-y)^2}\, .}}
 We first of all renormalize the upper vertex (connecting points
$z=0, u$, and $v$), by integrating by parts in $z$ and separating the
term integrated by parts into a trace and traceless piece (in the
indices $b$ and $c$). The surface term is finite by power counting, and
the trace piece easily renormalizes to a structure of the form
\eqn\someform{ \left(\g_d\g_\m\g_a\right){1\over (x-y)^2}\,\pa_a{1\over
x^2}\, \pa_d{\ln y^2M_V^2\over y^2}\, ,}
 which needs to be renormalized one more time (integration by parts
and
separation into trace and traceless pieces), giving a $\ln^2$ promotion
through the use of a DR identity.
This is easily done, and now we are finally left with the global
divergence ($x\sim y\sim 0$) present in the traceless piece (in $bc$) we
got after the first integration by parts. That term is power counting
log divergent, and one might think the traceless combination of
derivatives then makes it finite, but this is not so: the point is that
there are other free indices in that expression ($a$ and $d$).
The integration is indeed made finite by tracelessness in the $bc$
indices, because the $a$ and $d$ derivatives can be brought out of the
integral, but there remains a global divergence as $x\sim y\sim 0$,
because the expression is not traceless in {\it all} indices. Attempting
to separate the trace pieces from four free indices is hopeless because
we cannot even do the integral in $u$ and $v$, and so we resort to a
technique used to renormalize the nonplanar 3-loop 4-point diagram in
$\l\f^4$, valid for primitively divergent expressions in general.
Details can be found in \fjl , and we do not give them here. The idea
consists in writing the factor ${1\over (x-y)^2}$ in front as
\eqn\primit{{1\over (x-y)^2}=(x-y)^2{1\over (x-y)^4}=
{-(x-y)^2\over 4}\ \sq\, {\ln (x-y)^2M^2\over (x-y)^2}\, .}

One can verify that the simple substitution above, as is, suffices to
renormalize the global divergence we had (that is, that term will have a
well-defined Fourier transform). The only subtlety in applying the mass
derivative to the resulting renormalized amplitude lies in fact in this
globally divergent piece. $M\dd{M}$ on the term above is
proportional to $\d^{(4)}(x-y)$ and, like in \fjl , one must verify that
what multiplies this is a representation of $\d^{(4)}(y-z)$ when
$x\rightarrow y$. Adding up all contributions, the mass derivative then
reads:
\eqn\mdmti{ M\dd{M}\,\, V_\m^{(2i)}={\a\over 2\p}\,\, V_\m^{(1)}+
{5\a^2\over 16\p^2}\,\, V_\m^{(0)}.}

The next diagram is that of Fig. 2j, with bare value:
\eqn\vtj{\eqalign{ V_\m^{(2j){\rm bare}}(x,y)=&
{-(ie)^5\over (4\p^2)^6}
\left(\g_\r\g_a\g_\m\g_b\g_\n\g_c\g_\r\g_d\g_\n\right)\times
\cr &
\pa_a{1\over x^2}\int\dudv \pa_b{1\over u^2}{1\over (x-v)^2}
\pa_c{1\over (u-v)^2}\pa_d{1\over (v-y)^2}{1\over (u-y)^2}\, .}}

The first step in renormalizing this diagram is integration by parts of
the $\pa_d$ and $\pa_c$ derivatives, and separation into trace and
traceless parts. There will be two of each; the first trace piece is
easily renormalizable with standard DR identities, and the second one
involves the structure
\eqn\anstruc{ \left(\g_d\g_a\g_\m\right)\pa_a{1\over x^2}\,
\pa_d{1\over y^2}\, K(x,y)\, ,}
 where
\eqn\kxy{K(x,y)=\int\du{1\over (x-u)^2(y-u)^2u^2}\, .}
 This is renormalized one more time by integrating the derivatives by
parts onto the $K$-function and separating trace and traceless pieces.
For the first time, we encounter renormalization structures that do not
correspond to any lower loop diagrams, and because of consistency with
RG equations, these must cancel in the only other diagram left to
compute, Fig. 2l (the nonplanar diagram). We find that indeed they do.
The only other divergent pieces remaining are the ones traceless in $cb$
and $cd$, which can have a global divergence for the same reason as the
equivalent term in the previous diagram: the presence of extra indices.
These are primitively divergent, like in the previous diagram, and are
solved in exactly the same way. Naturally, the diagram identical to this
one, but with the vertex subdivergence on the opposite fermion leg,
Fig. 2k, is solved in the same fashion, and leads to the same
contribution to the 2-loop RG equations. We present the result of the
mass derivative on these diagrams after the inclusion of the nonplanar
diagram, Fig. 2l, precisely because there are structures in these three
diagrams which do not correspond to any lower loop diagram, and which
cancel when they are added. So, finally, the nonplanar diagram is:
\eqn\vtl{\eqalign{ V_\m^{(2l){\rm bare}}(x,y)=&
-{(ie)^5\over(4\p^2)^6}
\left(\g_\n\g_a\g_\r\g_b\g_\m\g_c\g_\n\g_d\g_\r\right)\times\cr &
\int\dudv \pa_a{1\over (x-u)^2}\pa_b{1\over u^2}\pa_c{1\over
v^2}\pa_d{1\over (v-y)^2}{1\over (x-v)^2(u-y)^2}\, .}}
 To renormalize this, we integrate $\pa_c$ by parts around $z=0$, and
again separate trace and traceless pieces. Within the trace pieces, we
will find the structure
\eqn\stst{ \g_\m\left[\dd{x^a}\left({1\over
x^2y^2}\dd{y^a}K(x,y)\right)
+{1\over x^2y^2}\dd{x^a}\dd{y^a}K(x,y)\right]\, ,}
and these will precisely cancel with structures found in the two
previous diagrams. We can now present the mass derivative acting on the
sum of these three diagrams:
\eqn\mdmtj{ M\dd{M}\,\, [V_\m^{(2j)}+V_\m^{(2k)}+V_\m^{(2l)}]
={\a\over\p}\,\, V_\m^{(1)}-
{\a^2\over 8\p^2}\,\, V_\m^{(0)}.}

We are now ready to consider RG equations. These are easy to verify once
we use the mass derivatives given above for the 2-loop vertices. The
2-loop $\G^{\psibar A\ps}_\m$ vertex function,
\eqn\gpsiaps{
\G_{\m}^{\psibar A\ps}(x,y)=ie\g_\m\d^{(4)}(x-y)\d^{(4)}(x)+
V_\m^{(1)}(x,y)+V_\m^{(2f+2g+2h+2i+2j+2k+2l)}(x,y),}
satisfies an RG equation, Eq.\rgtp ,
confirming all the results given previously for different RG
functions, and yielding also the 2-loop $\b$-function:
\eqn\betaa{\b (\a )={2\a^2\over 3\p}+{\a^3\over 2\p^2}\, .}
This matches the value gotten by a background field calculation on the
2-loop vacuum polarization.

\vskip 1pc

\subsect{2.4 Ward identity}

At two loops, the method employed previously to verify Ward
identities, viz., integrating over an external variable, becomes
computationally difficult and not very illuminating.
For our purposes here, we shall consider instead the following simpler
procedure: we apply the logarithmic mass derivative to both sides of the
(2-loop) Ward identity; as we have seen above, this yields 1-loop and
tree-level vertices
and self-energies, whose Ward identities have already been verified, so
that our problem is reduced to that of matching the coefficients of
1-loop and tree-level quantities on both sides of the identity.
We have, on the one hand, gained much in simplicity but, of course, on
the other hand, some information contained in the Ward identity will
thus be lost, namely, the contact terms coming from the finite parts of
vertices, since these latter have no mass scales and will vanish when
the mass derivative is applied. Although in this way we cannot derive
mass relations like we did at one loop, the match we find represents a
highly nontrivial consistency check of our 2-loop computations.

Adding up all the contributions from Eqs.\mdmstc , \mdmstd\ and \mdmste
, we find:
\eqn\mdmscde{ M\dd{M}\,\,\S^{(2c+2d+2e)}={\a\over 2\p}\,\,\S^{(1)}-
{7\a^2\over 16\p^2}\,\,\S^{(0)}}
and from Eqs.\mdmvtf , \mdmth , \mdmti , \mdmtj ,
\eqn\mdmvfghijkl{ M\dd{M}\,\, V_\m^{(2f+2g+2h+2i+2j+2k+2l)}={\a\over
2\p}\,\, V_\m^{(1)}-{3\a^2\over 16\p^2}\,\, V_\m^{(0)}-{2\a\over 3\p}
V_\m^{(1)}(a=0).}

Once we use the fact that, from Eq.\wiresult
\eqn\intv{\int d^4\!y \dd{z^\m}V^{(1)}_\m(x,y;a=0)={3\a\over 8\p}\,\,
\S^{(0)}(x)\,\, ,}
 we immediately find the amplitudes we have calculated above do
indeed
satisfy the 2-loop Ward identity. It is also worthwhile noting,
furthermore, that in fact we can divide the above 2-loop amplitudes
into three sets that separately verify the Ward identity. They are:
\eqn\sepwia{{\pa\over \pa z^\m}V_\m^{(2f+2g+2i)}(x,y)=
-ie[\d^{(4)}(x)-\d^{(4)}(y)]\S^{(2c)}(x-y)\, ,}
\eqn\sepwib{{\pa\over \pa z^\m}V_\m^{(2j+2k+2l)}(x,y)=
-ie[\d^{(4)}(x)-\d^{(4)}(y)]\S^{(2d)}(x-y)\, ,}
\eqn\sepwic{{\pa\over \pa z^\m}V_\m^{(2h)}(x,y)=
-ie[\d^{(4)}(x)-\d^{(4)}(y)]\S^{(2e)}(x-y)\, .}

The vertex diagrams in each one of these sets are generated by attaching
an external photon line in every possible way to an internal fermion
line of the corresponding self-energy diagram. This is a vestige of the
fact that each bare vertex -- individually -- formally satisfies a Ward
identity with the self-energy gotten by eliminating the external
photon line from that vertex diagram.


\newsec{Structured Renormalization Group}

In this section we point out a certain structure exhibited
by the renormalization of the different relevant vertex functions of
QED. For that purpose, we gather here the results of the mass derivative
of all 2-loop amplitudes.

\ni Vacuum polarization:
   \eqn\rgvac{
M\dd{M}\,\, [2\P_{\m\n}^{(2a)}+\P_{\m\n}^{(2b)}]=-{\a^2\over 2\p^2}
                 (\pa_\m\pa_\n -\d_{\m\n}\sq )\,\d (x) }
Self-energy:
\eqn\mmc{
M\dd{M}\,\,\S^{(2c)}=-{\a\over 2\p}\,\,\S^{(1)}-
                 {\a^2\over 16\p^2}\,\,\S^{(0)}  }
\eqn\mmd{
M\dd{M}\,\, \S^{(2d)}=\,\, {\a\over \p}\,\,\S^{(1)}-
                 {\a^2\over 8\p^2}\,\,\S^{(0)}   }
\eqn\mme{
M\dd{M}\,\,\S^{(2e)}=-{\a^2\over 4\p^2}\,\,\S^{(0)} }
Vertex:
\eqn\mmfg{
M\dd{M}\,\, [V_\m^{(2f)}+V_\m^{(2g)}]=-{\a\over\p}\,\, V_\m^{(1)}-
                 {3\a^2\over 8\p^2}\,\, V_\m^{(0)}    }
\eqn\mmh{
M\dd{M}\,\, V_\m^{(2h)}=-{2\a\over 3\p}\,\, V_\m^{(1)}(a=0)  }
\eqn\mmi{
M\dd{M}\,\, V_\m^{(2i)}=\,\, {\a\over 2\p}\,\, V_\m^{(1)}+
                 {5\a^2\over 16\p^2}\,\, V_\m^{(0)}  }
\eqn\mmjkl{
M\dd{M}\,\, [V_\m^{(2j)}+V_\m^{(2k)}+V_\m^{(2l)}]=\,\, {\a\over\p}\,\,
                 V_\m^{(1)}-{\a^2\over 8\p^2}\,\, V_\m^{(0)}  }

The feature the above equations clearly display, which has already been
alluded to in Section 2, is the absence of promotion of 1-loop
logarithms at two loops for $i)$ the 2-loop vacuum polarization,
Eq.\rgvac , $ii)$
the fermion self-energy with a vacuum polarization subdiagram, Eq.\mme ,
and
$iii)$ the vertex with a vacuum polarization subdiagram, Eq.\mmh .
All the rest
follow the expected pattern of log promotions. Given that the Ward
identities provide a relation between $M_\S$ and $M_V$, but not between
$M_\P$ and anything else, we might then understand the above as a
manifest, converse statement to the Ward identity, namely, that the
renormalization of the vacuum polarization runs entirely independently
of the renormalization of the self-energy and vertex. Thus, for the
2-loop vacuum polarization, although both $\P_{\m\n}^{(2a)}$ and
$\P_{\m\n}^{(2b)}$ contain promoted logs of $M_\S$ and $M_V$ (cf.
Eqs.\pmn\ and \pmntbx ) coming from their respective subdivergences,
in the entire 2-loop amplitude these promotions cancel. This is as it
should be, since otherwise RG equations would imply a relation between
the mass scales $M_\S$ (or $M_V$) and $M_\P$, which cannot happen if the
renormalization of these amplitudes is to be independent. For the
self-energy and vertex with vacuuum polarization subdivergences, the
same happens: no $\ln^2M_\P^2$ occur. The mass derivative of the vertex
does give a 1-loop vertex, but it is in Landau gauge, and thus finite.
For these amplitudes then, we can state that {\it there are no genuine
2-loop divergences}, and that is a direct consequence of the {\it lack}
of a Ward identity relating the vacuum polarization to any other vertex
functions.

We see then that the renormalization of these amplitudes has, so to
speak, split into different sectors.  A careful study of this structured
renormalization pattern should furthermore allow us to make predictions
about the renormalization at higher loops. In \itzu (cf. p.423), for
instance, the statement is made that all higher-loop vacuum polarization
amplitudes
with a single fermion loop (i.e., with no lower-loop vacuum polarization
subdivergences) do not contain genuine higher loop divergences
($\ln^2$'s, or $1/\e^2$'s in dimensional regularization, etc.). This
feature, which may seem fortuitous in other renormalization methods,
appears naturally in \dr .

\newsec{Chiral Anomaly}

In this section, we review the computation of the anomalous triangle
amplitude
$\langle j_\m (x)j_\n (y)j_\l^5(z)\rangle$  at one loop \fjl , and
carry it on
partially at two loops. Some of the main attractive features
of \dr\ are made manifest. We also compare it to another coordinate
space,
regulator-free computation of the anomaly due to K. Johnson \dan\ for
the sake of completeness.

Since differential renormalization is
strictly 4-dimensional and does not introduce any unphysical
regulator fields, one is able to avoid the complications introduced by
standard methods such as dimensional regularization (in necessitating
{\it ad hoc} $\g_5$-prescriptions away from $d=4$), and Pauli-Villars
regularization. Furthermore,
 \dr\ does not present the anomaly as coming from a symmetry
that is broken by regularization artifacts, as is usually the case.
Instead, we shall see that the bare triangle amplitude, without the
$\g$-matrix trace factor in front, has singularities which bring in two
renormalization scales and these, when the trace is included, lead to a
finite amplitude with a continuous one-parameter shift freedom given by
the quotient of these scales.
It turns out that this freedom cannot accomodate simultaneously
vector an axial conservation.
Therefore, the triangle amplitude is
overconstrained by the two Ward identities. In our scheme, both
vector and axial
symmetries appear manifestly on the same footing, and the shift in the
anomaly from one to the other is transparent.

We begin with the 1-loop computation. The basic elements have already
been spelled out in \fjl . Here we simply sketch the key points.
At one loop, the anomalous triangle is (Fig. 3):
\eqn\triangle{
T_{\m\n\l}(x,y)=\langle j_\m (x)j_\n (y)j_\l^5(0)\rangle
=2 \tr [\g_5\g_\l \g_a\g_\n\g_b\g_\m\g_c]\,\,
\pa_a{1\over y^2}\,\pa_b{1\over (x-y)^2}\,\pa_c{1\over x^2}\,\, , }
where the factor of 2 reflects the inclusion of the Bose symmetrized
diagram. The term
$$t_{abc}(x,y)=\pa_a{1\over y^2}\,\pa_b{1\over (x-y)^2}\,\pa_c{1\over
x^2}$$
 is power counting $L^{-9}$ and thus linearly divergent. This is
renormalized in the same way we have treated triangles previously, with
the difference that now {\it two} derivatives need to be taken out. This
has already been done in \fjl , and we present the final result:
$$ t_{abc}(x,y)=F_{abc}(x,y)+S_{abc}(x,y)\,\, .$$
$F_{abc}$ is the finite part
\eqn\finitef{ \eqalign{ F_{abc}(x,y)&=
{\pa^2\over\pa x_a\pa y_b}\ \left[ {1\over x^2y^2}\ {\pa_c}\
{1\over (x-y)^2}\right]
+{\pa\over\pa x_a}\left[ {1\over x^2y^2}\left( \pa_b\pa_c -
{\d_{bc}\over 4}
\sq\right){1\over (x-y)^2}\right] \cr &
-{\pa\over\pa y_b}\left[ {1\over x^2y^2}\left(\pa_a\pa_c-
{\d_{ac}\over 4}
\sq\right){1\over (x-y)^2}\right]
-{1\over x^2y^2 }\left[\pa_a\pa_b\pa_c -{1\over 6}\left(
\d_{(ab}\pa_{c)}\right)\sq\right]{1\over (x-y)^2}
\ , }}
where $()$ means unnormalized symmetrization in all indices,
and $S_{abc}$ is the renormalization piece:
\eqn\singulars{\eqalign{
S_{abc}(x,y)&= {1\over 4} \pi^2 \left\{ \left[ \delta_{bc}
{\partial\over\partial x_a} - \delta_{ac} {\partial\over\partial
y_b}\right] \delta(x-y) \sq {\ln M^2_1 x^2\over x^2} \right.\cr
-{1\over 3}&\left.\left[ \delta_{bc} \left( {\partial\over\partial x_a}
- {\partial\over\partial y_a}\right) +
\delta_{ac} \left( {\partial\over\partial x_b}
- {\partial\over\partial y_b} \right) + \delta_{ab} \left(
{\partial\over\partial x_c} - {\partial\over\partial y_c}\right)\right]
 \delta(x-y) \sq {\ln M^2_2 x^2\over x^2} \right\} \ .\cr }}
We note that, like for the 1-loop fermion self-energy, two
different mass scales
have been used to renormalize the different divergent trace pieces.
Not only are we entitled to do that, but in fact one more time it will
prove crucial that we do so.

At this point, we apply the traces outside to get $T_{\m\n\l}(x,y)$:
\eqn\tmnl{T_{\m\n\l}(x,y)=R_{\m\n\l}(x,y)+a_{\m\n\l}(x,y)\,\, ,}
\ni where $R_{\m\n\l}$ comes from the finite part $F_{abc}$, and
\eqn\amnl{a_{\m\n\l}(x,y)=-16\p^4\ln{M_1\over M_2}\,\
\epsilon_{\m\n\l\r}\left(
{\pa\over\pa x_\r}-{\pa\over\pa y_\r}\right)\d (x)\d (y)\, .}

It is fairly simple to see that for $x\ne y\ne 0$, $T_{\m\n\l}$ is
conserved on all channels. However, just as in the vertex WI studied
previously,
the subtleties are of course in the contact terms. It is also worthwhile
noting that while $S_{abc}$ (and thus $t_{abc}$) indeed contains
divergences, the precise $\g$-structure in front arranges things such as
to give the final combination $\ln{M_1\over M_2}$ as the only
renormalization
mass dependence, implying that $a_{\m\n\l}$ (and
thus $T_{\m\n\l}$) is
actually finite (because $M\dd{M}$ on that vanishes). This leads us then
to a very clear physical picture: we have a finite
anomalous triangle, which however contains an ambiguity -- in the choice
of $M$s -- coming from a power counting linearly divergent Feynman
diagram.

By Fourier transforming into momentum space, one can verify the
conservation laws satisfied by $R_{\m\n\l}$ on all three channels. Given
the form of $a_{\m\n\l}$ found above, the vector and axial WIs on
$T_{\m\n\l}$ then read:
\eqn\conserveqns{\eqalign{
\dd{x^\m}T_{\m\n\l}(x,y)&=8\p^4(1+2\ln M_1/M_2)
\e_{\n\l\m\r}{\pa\over\pa x_\m}{\pa\over\pa y_\r}\d (x)\d (y)\cr
\dd{y^\n}T_{\m\n\l}(x,y)&=8\p^4(1+2\ln M_1/M_2)
\e_{\m\l\n\r}{\pa\over\pa y_\n}{\pa\over\pa x_\r}\d (x)\d (y)\cr
-\left({\pa\over\pa x_\l}+{\pa\over\pa y_\l}
\right) T_{\m\n\l}(x,y)&=16\p^4(1-2\ln M_1/M_2)\e_{\m\n\l\r}
{\pa\over\pa x_\l}{\pa\over\pa y_\r}\d (x)\d (y)\,\, .}}
This is the final and, as it were, a most ``democratic" expression of
the anomaly: we can tune $M_1/M_2$ so that $T_{\m\n\l}$ is
conserved either on the vector channels or on the axial one, but never
on both; we have been able to complete the calculation in a
``scheme-free" fashion all the way to the end, without having to commit
at any point to conservation on a particular channel.

We now proceed to the 2-loop calculations. The relevant diagrams are
indicated in Fig. 4. Writing the amplitude as
\eqn\bla{T_{\m\n\l}^{(2)}(x,y)=A_{\m\n\l}(x,y)+B_{\m\n\l}(x,y)\,\, ,}
where $A_{\m\n\l}$ contains the contributions of the diagrams with
fermion self-energy insertions, and $B_{\m\n\l}$ contains the
contributions of the diagrams with the vertex insertions, we can
immediately write, as the result of a trivial computation:
\eqn\aamnl{\eqalign{  A_{\m\n\l}(x,y)=
{e^2\over 32\p^4}\tr& [\g_5\g_\l\g_a\g_\n\g_b\g_\m\g_c]  \left(
{\pa_a}{1\over y^2}\pa_b{1\over (y-x)^2}\pa_c{\ln x^2M_\S^2\over x^2}+
\right.\cr &\pa_a{\ln y^2M_\S^2\over y^2}\pa_b{1\over
(y-x)^2}\pa_c{1\over x^2}+
  \left.
\pa_a{1\over y^2}\pa_b{\ln (y-x)^2M_\S^2\over (y-x)^2}
\pa_c{1\over x^2}\right)\,\, .
   }}
To compute the diagrams related to vertex corrections we operate
as follows. Let us take, for instance, the correction to the vertex at
$y$, which leads to the following integral:
\eqn\qwe{\int d^4ud^4v {1\over (u-v)^2}\partial_a{1\over
u^2}\partial_b{1\over
(u-y)^2} \pa_c{1\over (y-v)^2}\partial_d{1\over (v-x)^2}\partial_e
{1\over x^2}  }
Treating the vertex in the standard way, we integrate $\pa_b$ by parts
over $y$, and separate a trace and a traceless piece, and a surface
term. The last two terms do not produce logarithms. It is only the trace
piece, which is very easy to compute, that brings in a log of the same
kind as in Eq.\aamnl.
Putting together the log pieces of the three vertex correction diagrams
we get
\eqn\bbmnl{\eqalign{  B_{\m\n\l}(x,y)=-
{e^2\over 32\p^4}\tr& [\g_5\g_\l\g_a\g_\n\g_b\g_\m\g_c]  \left(
{\pa_a}{1\over y^2}\pa_b{1\over (y-x)^2}\pa_c{\ln x^2M_V^2\over x^2}+
        \right.\cr&
\pa_a{\ln y^2M_V^2\over y^2}\pa_b{1\over (y-x)^2}\pa_c{1\over x^2}+
  \left.
\pa_a{1\over y^2}\pa_b{\ln (y-x)^2M_V^2\over (y-x)^2}
\pa_c{1\over x^2}\right)+...\,\, ,  }}
where the dots indicate the finite pieces not written explicitly here.
Clearly, $A_{\m\n\l}$ and $B_{\m\n\l}$ cancel exactly except for the
fact
that the mass scale involved in $A_{\m\n\l}$ is $M_\S$ whereas the one in
$B_{\m\n\l}$ is
$M_V$. The one-loop Ward identity shows that the mistmatch is
precisely proportional to the 1-loop amplitude, thus showing that the
entire 2-loop amplitude is finite. The finite parts of $B_{\m\n\l}$ are
extremely lengthy to calculate and we have not found a compact way to
perform the computation. In fact, the total amplitude is expected
to vanish identically since, otherwise, as we now explain, the anomaly
would get a finite renormalization.

QED is conformally invariant up to the photon propagator
and the appearance of renormalization scales. Thus, the anomalous
triangle is conformally covariant at one loop because it is finite and
contains no photon lines.
In any coordinate space treatment, this can be
verified explicitly to one loop, whereas in momentum space
calculations, this is obscured because plane waves transform easily
under translations but not under conformal tranformations.
This conformal property, when conjectured to all loops, is
powerful enough to show the vanishing of all higher-loop
triangle amplitudes, in an argument due to Baker and Johnson \baker .
{}From the study of conformal transformations on functions of three
variables, it turns out there is a unique nonlocal, conformal covariant,
parity-odd,
dimension 3, VVA tensor \schreier , and the 1-loop triangle has
precisely this
form. Now, from uniqueness, if the triangle amplitude is covariant to all
orders, it
means higher-loop contributions also have to have the form of the 1-loop
triangle. But that would mean a renormalization of the very
structure which gives rise to the chiral anomaly, and thus a
renormalization of the anomaly itself, which is forbidden by the
Adler-Bardeen theorem. Therefore, all higher-loop contributions to the
basic triangle must vanish identically.

Let us finally mention a very nice coordinate space computation of
the anomaly which makes no use of an ultraviolet regulator, due to
Johnson \dan . His starting point is the conformally covariant and
manifestly finite form of the triangle gotten by acting out the
$\g$-matrix trace in Eq.\triangle\ from the beginning.
Because there is,
apart from this nonlocal structure, a unique contact term with precisely
the same dimension and conformal properties \schreier , the triangle
amplitude can have this local ambiguity, and its coefficient can be
chosen to give conservation on either channel but not both. If we want
conservation on the $x$ channel, for instance, we have:
\eqn\kenano{\dd{x^\m}\left[
{T_{\m\n\l}(x,y)}+a\e^{\m\n\l\s}\left(\dd{x^\s}
-\dd{y^\s}\right)\d (x){\d (y)}\right]=0\,\, ,}
\ni where we have shown explicitly the contact term in question, with
coefficient $a$ to be determined. To find $a$, one then
integrates $\pa_\m T_{\m\n\l}$ in $x$ and $y$ against the ``test"
function $x^\a y^\b$. This integral, remarkably, is entirely determined
from the long-distance behavior of $T_{\m\n\l}$, and the only cutoff
needed to perform it is an {\it infrared} one. Johnson also points
out that
this situation is reminiscent of the Poisson equation in classical
electrostatics, where the coefficient of a contact term in a
differential equation (the charge of a point particle) is also
determined by the long-distance behavior of a field (Gauss' theorem).

\newsec{Conclusion}

In this paper, we have presented a detailed analysis of the
differential renormalization of QED up to two loops. All computations
are reasonably simple when compared to other approaches.

Our results can be summarized as follows. We have found explicit
expressions for the 1- and 2-loop renormalized amplitudes of the vacuum
polarization,
self-energy of the fermion and vertex. Only some finite parts of the
latter are left in the form of integrals. Ward Identities are verified
and provide independent checks of the computations. Furthermore, the
amplitudes obey renormalization group equations which yield
the beta function and the various anomalous dimensions
of all the basic fields in the theory. This renormalization group
equations display a natural organization due to the freedom to use
different renormalization subtractions for quantities not related
 by WIs.
We have also presented the treatment of the anomalous triangle which,
in the very spirit of \dr , is regulator-free. It sides with other
techniques which are based on the interplay between potential
counterterms and symmetries rather than on explicit computations on
a regulated theory.

 One is rapidly enticed by the ease of computation of \dr . It
is our
opinion that the method also deserves further consideration because of
its natural treatment of chiral problems.

\vskip 1pc

\subsect{Acknowledgments}

We thank D.Z. Freedman warmly for his active participation in
different stages of this paper and K. Johnson for his insight on
anomalies. We
would also like to acknowledge discussions with X. Vilas\iii s-Cardona,
and a critical reading of the manuscript by R. Tarrach.
This work was supported in part by CAICYT grant \# AEN90-0033, and by
EEC Science Twinning grant \# SCI-000337. One of us (PH) also
acknowledges a grant from MEC, Spain.

\vskip 2pc

\appendix{A}{Basic Differential Renormalization Identities}

The following are the DR identities used to renormalize all of the
amplitudes presented in the text:

\bigskip
\eqn\john{
{1\over x^4}= -{1\over 4} \sq {\ln x^2M^2\over x^2} }
\eqn\paul{
{1\over x^6} = -{1\over 32}
\sq\sq{\ln x^2M^2\over x^2} }
\eqn\mary{
{\ln x^2M^2\over x^4} = -{1\over 8} \sq {  \ln^2(x^2M^2)+2
\ln x^2M'^2\over  x^2} }

\bigskip

\appendix{B}{Two-loop renormalized vertices}

We present here the final expressions for the 2-loop renormalized
vertices. In the expressions below, a
derivative w.r.t. $z^\m$ means $\dd{z^\m}=-\dd{x^\m}-\dd{y^\m}$.

\eqn\apf{\eqalign{
&V_\m^{(2f)}(x,y)=\cr
 &-{i\over 16\p^3}\left({\a\over\p}\right)^{5/2}
\left(\g_b\g_\m\g_a\right)\left\{
{\pa\over\pa z^b}\left[
{\ln y^2M^2\over y^2 (x-y)^2}
\pa_a{1\over x^2}\right]+\right.\cr
&\left.
{\ln y^2M^2\over y^2 (x-y)^2}
\left(\pa_a\pa_b -{\d_{ab}\over
4}\sq\right){1\over x^2}
+{\p^2\over 8}\d_{ab}\d^{(4)}(x)
\sq\, {\ln^2y^2M_{\S}^2+2\ln y^2M^2\over y^2}\right\}.}}

\eqn\aph{\eqalign{
&V_\m^{(2h)}(x,y)=\cr
&{i\over 96\p^3}\left({\a\over\p}\right)^{5/2}
\left\{\left(\g_\r\g_a\g_\m\g_b\g_\s\right)
\left[-{\pa\over\pa x^a}\left({1\over x^2}\pa_b{1\over y^2}
(\pa_\r\pa_\s -\d_{\r\s}\sq )L(x-y)\right)+\right.\right.\cr &
\left.\left.
{\pa\over\pa y^b}\left({1\over x^2 y^2}\pa_a
(\pa_\r\pa_\s -\d_{\r\s}\sq )L(x-y)\right)\right]
-{16\over x^2y^2}\g_a\left(\pa_a\pa_\m
-{\d_{a\m}\over 4}\sq\right){\ln (x-y)^2M_\P^2\over
(x-y)^2}\right\}\, ,}}
 where $L(x)=\ln x^2\m^2-{1\over 2}\ln^2x^2M_\P^2$.

\eqn\api{\eqalign{
&V_\m^{(2i)}(x,y)=\cr
&{i\over 32\p^7}\left({\a\over\p}\right)^{5/2}
\int\dudv\left\{\left(\g_d\g_b\g_\m\g_c\g_a\right)
\dd{z^c}\left[{1\over (x-y)^2 v^2}\pa_b{1\over u^2}{1\over
(u-v)^2}
\pa_a{1\over (x-u)^2}\pa_d{1\over (v-y)^2}\right]\right.\cr
&\left. -\left(\g_d\g_c\g_a\right){(x-y)^2\over 2}\sq\,
{\ln (x-y)^2M^2\over (x-y)^2}{1\over v^2 (u-v)^2}
(\pa_c\pa_\m -{\d_{c\m}\over 4}\sq ){1\over u^2}\pa_a
{1\over (x-u)^2}\pa_d{1\over (v-y)^2}
\right\} +\cr &
{i\over 16\p^3}\left({\a\over\p}\right)^{5/2}
\left(\g_d\g_\m\g_a\right)\left\{
\dd{z^d}\left[{1\over (x-y)^2}\pa_a
{\ln y^2M^2\over y^2 x^2}\right]+{1\over (x-y)^2}
(\pa_a\pa_d -{\d_{ad}\over 4}\sq ){1\over x^2}
{\ln y^2M^2\over y^2}\right\}\cr &
-{i\over 64\p}\left({\a\over\p}\right)^{5/2}\g_\m
\d^{(4)}(x)
\sq\, {\ln^2y^2M_{V}^2+2\ln y^2M^2\over y^2}\, .}}

\eqn\apj{\eqalign{& V_\m^{(2j)}(x,y)=\cr &{-i\over 64\p^7}
\left({\a\over\p}\right)^{5/2}
\left(\g_\r\g_a\g_\m\g_b\g_d\g_\r\g_c\right)\int\dudv\left\{\dd{z^d}
\left[\pa_a{1\over x^2} \pa_b{1\over u^2}{1\over (x-v)^2}
\pa_c{1\over (u-v)^2}{1\over (v-y)^2}{1\over (u-y)^2}\right]\right.\cr
&-
\left.\pa_a{1\over x^2}\left[
\pa_b{1\over u^2}(\pa_c\pa_d-{\d_{cd}\over 4}\sq ){1\over (u-y)^2}+
\pa_d{1\over (u-y)^2}
(\pa_c\pa_b-{\d_{cb}\over 4}\sq){1\over u^2}\right]
{1\over (x-v)^2(u-v)^2(v-y)^2}\right\} \cr
&+
{i\over 16\p^3}\left({\a\over\p}\right)^{5/2}
\left(\g_b\g_\m\g_a\right)\left\{
\dd{z^a}\left[{1\over x^2}\pa_b{1\over y^2}
{\ln (x-y)^2M^2\over (x-y)^2}\right]+{1\over x^2}
(\pa_a\pa_b -{\d_{ab}\over 4}\sq ){1\over y^2}
{\ln (x-y)^2M^2\over (x-y)^2}\right\}   \cr
&-
{i\over 64\p}\left({\a\over\p}\right)^{5/2}\g_\m
\d^{(4)}(y)\sq\, {\ln^2(x-y)^2M_V^2+2\ln (x-y)^2M^2\over (x-y)^2}+
{i\over 16\p^5}\left({\a\over\p}\right)^{5/2}
\left(\g_b\g_a\g_\m\right)\left\{
\dd{x^a}\left[{1\over x^2}\pa_b{1\over y^2}K(x,y)\right]\right.\cr
&+
\left.\dd{y^b}\left[{1\over x^2y^2}\dd{x^a}K(x,y)\right]-{1\over x^2y^2}
\left(\dd{x^a}\dd{y^b}-{\d_{ab}\over 4}\sq\,\right)K(x,y)
\right\}\cr
&-
{i\over 32\p^3}\left({\a\over\p}\right)^{5/2}\g_\m
\left\{ {1\over 4}\sq\,{\ln x^2M^2\over x^2}\sq\,{\ln y^2M^2\over y^2}+
\dd{x^a}\left[{1\over (x-y)^2y^2}\lrhup{{\pa\over\pa x^a}} {\ln
x^2M^2\over x^2}\right] \right.         \cr
&+\left.
\p^2
\d^{(4)}(x-y)\sq\, {\ln^2x^2M_{V}^2+2\ln x^2M^2\over x^2}+
\dd{y^a}\left[{1\over (x-y)^2x^2}\lrhup{{\pa\over\pa y^a}}{\ln
y^2M^2\over y^2}\right]
\right\}\, .}}

\eqn\apl{\eqalign{& V_\m^{(2l)}(x,y)=\cr &{i\over 128\p^7}
\left({\a\over\p}\right)^{5/2}
\left(\g_\n\g_a\g_\r\g_b\g_\m\g_c\g_\n\g_d\g_\r\right)
\int\dudv\ \dd{z^c}
\left[\pa_a{1\over (x-u)^2} \pa_b{1\over u^2}{1\over v^2}
\pa_d{1\over (v-y)^2}{1\over (x-v)^2(u-y)^2}\right]\cr
&
-{i\over 8\p^7}\left({\a\over\p}\right)^{5/2}\g_b
\int\dudv\ \pa_a{1\over (x-u)^2}(\pa_b\pa_\m-{1\over
4}\d_{b\m}\sq){1\over u^2}\pa_a{1\over (v-y)^2}\cr
&
+{i\over 8\p^5}\left({\a\over\p}\right)^{5/2}\g_\m\dd{x^a}
\left[{1\over x^2y^2}\dd{y^a}K(x,y)\right]+
{i\over 16\p^3}\left({\a\over\p}\right)^{5/2}\g_\m\left\{
{1\over 4}\sq\,{\ln x^2M^2\over x^2}\sq\,{\ln y^2M^2\over y^2}+
\right.\cr
& \left.
-{\partial\over \partial z^a}\left[{1\over
(x-y)^2y^2}\lrhup{{\pa\over\pa x^a}} {\ln x^2M^2\over x^2}\right]+
\p^2
\d^{(4)}(x-y)\sq\, {\ln^2x^2M_{V}^2+2\ln x^2M^2\over x^2}
\right\}\, .}}
\bigskip

Figure captions:

Figure 1: 1-loop 1PI diagrams of QED.

Figure 2: 2-loop diagrams of QED.

Figure 3: 1-loop anomalous triangle diagram.

Figure 4: 2-loop contributions to anomalous ABJ amplitude.
\eject \listrefs \end

\eject

\cl{{\bf Figures}}
\vskip 10pc

\cl{Figure 1: 1-loop 1PI diagrams of QED.}\vskip 25pc

\vfill
\cl{Figure 2: 2-loop diagrams of QED.}\eject

\vskip 18pc

\cl{Figure 3: 1-loop anomalous triangle diagram.}

\vskip 28pc

\cl{Figure 4: 2-loop contributions to anomalous ABJ amplitude.}
\eject
\listrefs
              \end